\documentclass[aps,prl,preprint,superscriptaddress,showpacs,preprintnumbers,amsmath,amssymb]{revtex4-1}

\usepackage{xspace}
\usepackage{graphicx} 
\usepackage{dcolumn}  
\usepackage[T1]{fontenc} 

\usepackage[english]{babel}
\graphicspath{{ps}}

\RequirePackage{xspace}
\RequirePackage[separate-uncertainty = true]{siunitx}
\RequirePackage{xfrac}

\DeclareSIUnit\lightc{\ensuremath{\mathit{c}}}
\newcommand{\mevcc}{\si{\MeV/\lightc\squared}\xspace}

\newcommand{\gevcc}{\si{\GeV/\lightc\squared}\xspace}
\newcommand{\gevc}{\si{\GeV/\lightc}\xspace}

\newcommand{\Br}{\ensuremath{\mathcal{B}}\xspace}
\newcommand{\Acp}{\ensuremath{\mathcal{A}_{CP}}\xspace}

\newcommand{\dz}{\ensuremath{D{}^0}\xspace}
\newcommand{\adz}{\ensuremath{\overline{D}{}^0}\xspace}
\newcommand{\dsp}{\ensuremath{D^{*+}}\xspace}

\newcommand{\aksz}{\ensuremath{\overline{K}{}^{*0}}\xspace}
\newcommand{\rhoz}{\ensuremath{\rho{}^{0}}\xspace}

\newcommand{\dztorg}{\ensuremath{\dz \to \rhoz \gamma}\xspace}

\newcommand{\dztopg}{\ensuremath{\dz \to \phi \gamma}\xspace}

\newcommand{\brdztorg}{\ensuremath{\Br\left(\dz \to \rhoz \gamma\right)}\xspace}
\newcommand{\brdztokg}{\ensuremath{\Br\left(\dz \to \aksz \gamma\right)}\xspace}
\newcommand{\brdztopg}{\ensuremath{\Br \left(\dz \to \phi \gamma\right)}\xspace}

\newcommand{\acpdztorg}{\ensuremath{\Acp\left(\dz \to \rhoz \gamma\right)}\xspace}
\newcommand{\acpdztokg}{\ensuremath{\Acp\left(\dz \to \aksz \gamma\right)}\xspace}
\newcommand{\acpdztopg}{\ensuremath{\Acp\left(\dz \to \phi \gamma\right)}\xspace}

\newcommand{\lumi}{\SI{943}{~\femto\barn^{-1}}}
\usepackage{relsize}
\def\babar{\mbox{\slshape B\kern-0.1em{\smaller A}\kern-0.1em
    B\kern-0.1em{\smaller A\kern-0.2em R}}}

\begin{document}

\vspace*{-3\baselineskip}
\resizebox{!}{3cm}{\includegraphics{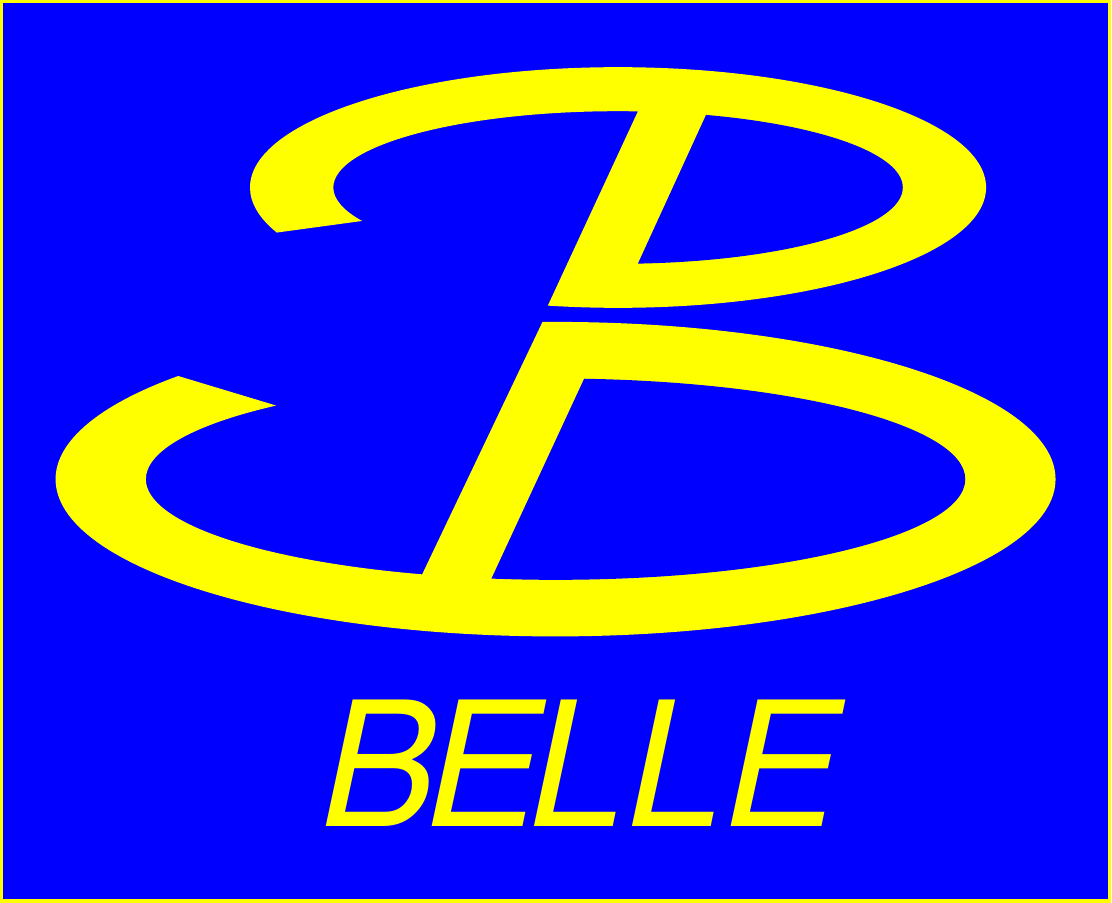}}

\preprint{\vbox{ \hbox{Belle Preprint 2016-11}
    \hbox{KEK Preprint 2016-24}
}}

\title{\boldmath Observation of $D^0\to \rho^0\gamma$ and search for $CP$ violation in radiative charm decays}

\noaffiliation
\affiliation{Aligarh Muslim University, Aligarh 202002}
\affiliation{University of the Basque Country UPV/EHU, 48080 Bilbao}
\affiliation{Beihang University, Beijing 100191}
\affiliation{University of Bonn, 53115 Bonn}
\affiliation{Budker Institute of Nuclear Physics SB RAS, Novosibirsk 630090}
\affiliation{Faculty of Mathematics and Physics, Charles University, 121 16 Prague}
\affiliation{Chonnam National University, Kwangju 660-701}
\affiliation{University of Cincinnati, Cincinnati, Ohio 45221}
\affiliation{Deutsches Elektronen--Synchrotron, 22607 Hamburg}
\affiliation{University of Florida, Gainesville, Florida 32611}
\affiliation{Justus-Liebig-Universit\"at Gie\ss{}en, 35392 Gie\ss{}en}
\affiliation{SOKENDAI (The Graduate University for Advanced Studies), Hayama 240-0193}
\affiliation{Gyeongsang National University, Chinju 660-701}
\affiliation{Hanyang University, Seoul 133-791}
\affiliation{University of Hawaii, Honolulu, Hawaii 96822}
\affiliation{High Energy Accelerator Research Organization (KEK), Tsukuba 305-0801}
\affiliation{J-PARC Branch, KEK Theory Center, High Energy Accelerator Research Organization (KEK), Tsukuba 305-0801}
\affiliation{IKERBASQUE, Basque Foundation for Science, 48013 Bilbao}
\affiliation{Indian Institute of Science Education and Research Mohali, SAS Nagar, 140306}
\affiliation{Indian Institute of Technology Bhubaneswar, Satya Nagar 751007}
\affiliation{Indian Institute of Technology Guwahati, Assam 781039}
\affiliation{Indian Institute of Technology Madras, Chennai 600036}
\affiliation{Indiana University, Bloomington, Indiana 47408}
\affiliation{Institute of High Energy Physics, Chinese Academy of Sciences, Beijing 100049}
\affiliation{Institute of High Energy Physics, Vienna 1050}
\affiliation{INFN - Sezione di Torino, 10125 Torino}
\affiliation{J. Stefan Institute, 1000 Ljubljana}
\affiliation{Kanagawa University, Yokohama 221-8686}
\affiliation{Institut f\"ur Experimentelle Kernphysik, Karlsruher Institut f\"ur Technologie, 76131 Karlsruhe}
\affiliation{Kennesaw State University, Kennesaw, Georgia 30144}
\affiliation{King Abdulaziz City for Science and Technology, Riyadh 11442}
\affiliation{Department of Physics, Faculty of Science, King Abdulaziz University, Jeddah 21589}
\affiliation{Korea Institute of Science and Technology Information, Daejeon 305-806}
\affiliation{Korea University, Seoul 136-713}
\affiliation{Kyungpook National University, Daegu 702-701}
\affiliation{\'Ecole Polytechnique F\'ed\'erale de Lausanne (EPFL), Lausanne 1015}
\affiliation{P.N. Lebedev Physical Institute of the Russian Academy of Sciences, Moscow 119991}
\affiliation{Faculty of Mathematics and Physics, University of Ljubljana, 1000 Ljubljana}
\affiliation{Ludwig Maximilians University, 80539 Munich}
\affiliation{Luther College, Decorah, Iowa 52101}
\affiliation{University of Maribor, 2000 Maribor}
\affiliation{Max-Planck-Institut f\"ur Physik, 80805 M\"unchen}
\affiliation{School of Physics, University of Melbourne, Victoria 3010}
\affiliation{University of Miyazaki, Miyazaki 889-2192}
\affiliation{Moscow Physical Engineering Institute, Moscow 115409}
\affiliation{Moscow Institute of Physics and Technology, Moscow Region 141700}
\affiliation{Graduate School of Science, Nagoya University, Nagoya 464-8602}
\affiliation{Kobayashi-Maskawa Institute, Nagoya University, Nagoya 464-8602}
\affiliation{Nara Women's University, Nara 630-8506}
\affiliation{National Central University, Chung-li 32054}
\affiliation{National United University, Miao Li 36003}
\affiliation{Department of Physics, National Taiwan University, Taipei 10617}
\affiliation{H. Niewodniczanski Institute of Nuclear Physics, Krakow 31-342}
\affiliation{Nippon Dental University, Niigata 951-8580}
\affiliation{Niigata University, Niigata 950-2181}
\affiliation{University of Nova Gorica, 5000 Nova Gorica}
\affiliation{Novosibirsk State University, Novosibirsk 630090}
\affiliation{Pacific Northwest National Laboratory, Richland, Washington 99352}
\affiliation{University of Pittsburgh, Pittsburgh, Pennsylvania 15260}
\affiliation{Theoretical Research Division, Nishina Center, RIKEN, Saitama 351-0198}
\affiliation{University of Science and Technology of China, Hefei 230026}
\affiliation{Showa Pharmaceutical University, Tokyo 194-8543}
\affiliation{Soongsil University, Seoul 156-743}
\affiliation{Stefan Meyer Institute for Subatomic Physics, Vienna 1090}
\affiliation{Sungkyunkwan University, Suwon 440-746}
\affiliation{School of Physics, University of Sydney, New South Wales 2006}
\affiliation{Department of Physics, Faculty of Science, University of Tabuk, Tabuk 71451}
\affiliation{Tata Institute of Fundamental Research, Mumbai 400005}
\affiliation{Department of Physics, Technische Universit\"at M\"unchen, 85748 Garching}
\affiliation{Toho University, Funabashi 274-8510}
\affiliation{Department of Physics, Tohoku University, Sendai 980-8578}
\affiliation{Earthquake Research Institute, University of Tokyo, Tokyo 113-0032}
\affiliation{Department of Physics, University of Tokyo, Tokyo 113-0033}
\affiliation{Tokyo Institute of Technology, Tokyo 152-8550}
\affiliation{Tokyo Metropolitan University, Tokyo 192-0397}
\affiliation{University of Torino, 10124 Torino}
\affiliation{Virginia Polytechnic Institute and State University, Blacksburg, Virginia 24061}
\affiliation{Wayne State University, Detroit, Michigan 48202}
\affiliation{Yamagata University, Yamagata 990-8560}
\affiliation{Yonsei University, Seoul 120-749}
  \author{T.~Nanut}\affiliation{J. Stefan Institute, 1000 Ljubljana} 
  \author{A.~Zupanc}\affiliation{Faculty of Mathematics and Physics, University of Ljubljana, 1000 Ljubljana}\affiliation{J. Stefan Institute, 1000 Ljubljana} 
  \author{I.~Adachi}\affiliation{High Energy Accelerator Research Organization (KEK), Tsukuba 305-0801}\affiliation{SOKENDAI (The Graduate University for Advanced Studies), Hayama 240-0193} 
  \author{H.~Aihara}\affiliation{Department of Physics, University of Tokyo, Tokyo 113-0033} 
  \author{S.~Al~Said}\affiliation{Department of Physics, Faculty of Science, University of Tabuk, Tabuk 71451}\affiliation{Department of Physics, Faculty of Science, King Abdulaziz University, Jeddah 21589} 
  \author{D.~M.~Asner}\affiliation{Pacific Northwest National Laboratory, Richland, Washington 99352} 
  \author{V.~Aulchenko}\affiliation{Budker Institute of Nuclear Physics SB RAS, Novosibirsk 630090}\affiliation{Novosibirsk State University, Novosibirsk 630090} 
  \author{T.~Aushev}\affiliation{Moscow Institute of Physics and Technology, Moscow Region 141700} 
  \author{R.~Ayad}\affiliation{Department of Physics, Faculty of Science, University of Tabuk, Tabuk 71451} 
  \author{V.~Babu}\affiliation{Tata Institute of Fundamental Research, Mumbai 400005} 
% ...
\collaboration{The Belle Collaboration}

\begin{abstract}
We report the first observation of the radiative charm decay $\dztorg$ and the 
first search for $CP$ violation in decays $\dztorg$, $\phi\gamma$, and $\aksz(892)
\gamma$, using a data sample of \lumi\ collected with the Belle detector at 
the KEKB asymmetric-energy $e^+e^-$ collider. The branching fraction is 
measured to be $\brdztorg=(1.77 \pm 0.30 \pm 0.07) \times 10^{-5}$, where the 
first uncertainty is statistical and the second is systematic. The obtained 
$CP$ asymmetries, $\acpdztorg=+0.056 \pm 0.152 \pm 0.006$, 
$\acpdztopg=-0.094 \pm 0.066 \pm 0.001$,  and 
$\acpdztokg=-0.003 \pm 0.020 \pm 0.000$, are consistent with no $CP$ 
violation. We also present an improved measurement of the branching fractions 
$\brdztopg=(2.76 \pm 0.19 \pm 0.10) \times 10^{-5}$ and 
$\brdztokg=(4.66 \pm 0.21 \pm 0.21) \times 10^{-4}$.
\end{abstract}

\pacs{11.30.Er, 13.20.Fc, 13.25.Ft}

\maketitle

{\renewcommand{\thefootnote}{\fnsymbol{footnote}}}
\setcounter{footnote}{0}
Within the Standard Model (SM), charge-parity ($CP$) violation in weak decays of hadrons arises due to a single irreducible phase in the Cabibbo-Kobayashi-Maskawa matrix~\cite{Kobayashi:1973fv} and is expected to be very small for charmed hadrons: up to a few $10^{-3}$~\cite{Bigi:2011re, Isidori:2011qw, Brod:2011re}.  
Observation of $CP$ violation above the SM expectation would be an indication of new physics. This phenomenon in the charm sector has been extensively probed
in the past decade in many different decays~\cite{Amhis:2014hma}, reaching a sensitivity below 0.1\% in some cases~\cite{Aaij:2016cfh}. 
The search for $CP$ violation in radiative charm decays is complementary to the searches that have been exclusively performed in hadronic or leptonic decays. 
Theoretical calculations~\cite{Isidori:2012yx, Lyon:2012fk} show that, in SM extensions with chromomagnetic dipole operators, sizable $CP$ asymmetries can be expected 
in $\dztopg$ and $\rhoz\gamma$ decays. No experimental results exist to date regarding $CP$ violation in any of the radiative $D$ decays.

Radiative charm decays are dominated by long-range non-perturbative processes that can enhance the branching fractions up to $10^{-4}$, whereas short-range interactions
are predicted to yield rates at the level of $10^{-8}$~\cite{Burdman:1995te, Fajfer:2015zea}. Measurements of branching fractions of these decays can therefore be used 
to test the QCD-based calculations of long-distance dynamics. The radiative decay \dztopg was first observed by Belle~\cite{Abe:2003yv} and later measured with 
increased precision by \babar{}~\cite{Aubert:2008ai}. In the same study, \babar{} made the observation of $D^0 \to \aksz(892) \gamma$. As for \dztorg, CLEO~II has set an upper limit on 
its branching fraction at $2\times10^{-4}$~\cite{Asner:1998mv}. 

In this Letter, we present the first observation of $\dztorg$, improved branching fraction measurements of $D^0\to \phi\gamma$ and $\aksz \gamma$, as well as the first
search for $CP$ violation in all three decays. Inclusion of charge-conjugate modes is implied unless noted otherwise. The measurements are based on \lumi\ of data 
collected at or near the $\Upsilon(nS)$ resonances ($n = 2,3,4,5$) with the Belle detector~\cite{Abashian:2000cg, Brodzicka:2012jm}, operating at the KEKB 
asymmetric-energy $e^+e^-$ collider~\cite{Kurokawa:2001nw,Abe:2013kxa}. The detector components relevant for our study are: a tracking system comprising a silicon
vertex detector and a 50-layer central drift chamber (CDC), a particle identification (PID) system that consists of a barrel-like arrangement of time-of-flight
scintillation counters (TOF) and an array of aerogel threshold Cherenkov counters (ACC), and a CsI(Tl) crystal-based electromagnetic calorimeter (ECL). All are 
located inside a superconducting solenoid coil that provides a 1.5 T magnetic field.

We use Monte Carlo (MC) events, generated using EVTGEN~\cite{Lange:2001uf}, JETSET~\cite{Sjostrand:1993yb} and PHOTOS~\cite{Golonka:2005pn}, followed with a GEANT3~\cite{Brun:1987ma} based detector simulation, representing six times the data luminosity, to devise selection criteria and investigate possible sources of background. 
The selection optimization is performed by maximizing $S/\sqrt{S+B}$, where $S$ ($B$) is the number of signal (background) events in a signal window of the reconstructed \dz invariant mass $\SI{1.8}{\gevcc}<M(\dz)<\SI{1.9}{\gevcc}$. The branching fraction of 
\dztorg is set to \num{3e-5} in simulations in accordance with Ref.~\cite{Isidori:2012yx}, while the branching fractions of the other two decay modes are set 
to their world-average values~\cite{Agashe:2014kda}.

We reconstruct $\dz$ mesons by combining a $\rho^0$, $\phi$, or a $\aksz$ with a photon. The vector resonances are formed from 
$\pi^+\pi^-$ ($\rho^0$), $K^+K^-$ ($\phi$), and $K^-\pi^+$ ($\aksz$) combinations. Charged particles are reconstructed in the tracking system. A likelihood ratio 
for a given track to be a kaon or pion is obtained by utilizing specific ionization in the CDC, light yield from the ACC, and information from the TOF. Photons 
are detected with the ECL and required to have energies of at least 540 MeV.  To suppress events with two daughter photons from a $\pi^0$ decay forming a 
merged cluster, we restrict the ratio of the energy deposited in a $3\times3$ array of ECL crystals ($E_{9}$) and that in the enclosing 
$5\times5$ array ($E_{25}$) to be above 0.94. About 63\% of merged clusters are rejected by this requirement. We retain candidate $\rhoz$, $\phi$, or $\aksz$ resonances if their invariant masses are 
within \SI{150}, \SI{11}, or \SI{60}{\mevcc} of their nominal masses~\cite{Agashe:2014kda}, respectively. The \dz mesons are required to originate 
from $\dsp \to \dz \pi{}^+$ in order to identify the \dz flavor and to suppress the combinatorial background. The associated track must satisfy the 
aforementioned pion-hypothesis requirement. The $D^0$ daughters are refitted to a common vertex, and the resulting $\dz$ and the slow pion candidate 
from $\dsp$ decay are constrained to originate from a common point within the interaction point region. Confidence levels exceeding $10^{-3}$ are 
required for both fits. To suppress combinatorial background, we restrict the energy released in the decay, $q\equiv M(\dsp) - M(\dz) - m(\pi^+)$, 
where $m$ is the nominal mass, to lie in a \SI{\pm0.6}{\mevcc} window around the nominal 
value~\cite{Agashe:2014kda}. To further reduce the combinatorial background contribution, we require the momentum of the \dsp in the center-of-mass system 
[$p_{\mathrm{CMS}}(\dsp)$] to exceed $2.72$, $2.42$, and $2.17$ \gevc in the $\rho^0\gamma$, $\phi\gamma$, and  $\aksz\gamma$ modes, respectively.

We measure the branching fractions and $CP$ asymmetries of aforementioned radiative decays relative to well-measured hadronic $D^0$ decays to 
$\pi^+\pi^-$, $K^+K^-$, and $K^-\pi^+$ for the \rhoz, $\phi$, and $\aksz$ mode, respectively. The signal branching fraction is
\begin{equation}
\Br_\mathrm{sig}=\Br_\mathrm{norm}\times \frac{N_\mathrm{sig}}{N_\mathrm{norm}}\times \frac{\varepsilon_\mathrm{norm}}{\varepsilon_\mathrm{sig}} \quad,
\end{equation}
where $N$ is the extracted yield, $\varepsilon$ the reconstruction efficiency, and \Br the branching fraction for the corresponding mode. The raw asymmetry 
in decays of $D^0$ mesons to a specific final state $f$,
\begin{equation}
 A_{\mathrm{raw}}=\frac{N(\dz\to f) - N(\adz\to \overline{f})}{N(\dz\to f)+N(\adz\to \overline{f})}, \label{eq:arawdef}
 \end{equation}
depends not only on the $CP$ asymmetry, $\Acp =[\Br(D^0\to f) - \Br(\overline{D}{}^0\to \overline{f})]/[\Br(D^0\to f) + \Br(\overline{D}{}^0\to \overline{f})]$, 
but also on the contributions from the forward-backward production asymmetry ($A_\mathrm{FB}$)~\cite{Berends:1973fd, Brown:1973ji, Cashmore:1985vp} and the 
asymmetry due to different reconstruction efficiencies for positively and negatively charged particles 
($A_\varepsilon^{\pm}$): $A_\mathrm{raw} = \Acp+A_{\mathrm{FB}}+A_\varepsilon^{\pm}$. Here, we have used a linear approximation assuming all terms to be small. The last two terms can be eliminated using the same normalization mode 
as used in the branching fraction measurements:
\begin{equation}
 \Acp^\mathrm{sig}=A^\mathrm{sig}_\mathrm{raw} - A^\mathrm{norm}_\mathrm{raw} + \Acp^\mathrm{norm}, \label{eq:acpcalc} 
\end{equation}
where $\Acp^\mathrm{norm}$ is the nominal value of $CP$ asymmetry of the normalization mode~\cite{Amhis:2014hma}.

The dominant background arises from $D^0 \to f^+f^- \pi^0$ decays, with the $\pi^0$ subsequently decaying to a pair of photons, e.g., $D^0 \to \phi \pi^0 (\to \gamma \gamma)$. 
If one of the daughter photons is missed in the reconstruction, the final state mimics the signal decay. Such events are suppressed with a dedicated $\pi^0$ veto 
in the form of a neural network~\cite{Feindt2006190}  constructed from two mass-veto variables, described below. The signal photon is paired for the first (second) 
time with all other photons in the event having an energy greater than 30 (75) MeV. The pair in each set whose diphoton invariant mass lies closest to 
$m(\pi^0)$ is fed to the network. The final criterion on the veto variable rejects about \SI{60}{\percent} of background while retaining \SI{85}{\percent} 
of signal. With this method, we reject 13\% more background at the same signal efficiency as compared to the veto used in previous 
Belle analyses~\cite{Koppenburg:2004fz}. A similar veto is considered for background from $\eta\to \gamma \gamma$, but is found to be ineffective due to 
the larger $\eta$ mass, which shifts the background further away from the signal peak.

We extract the signal yield and $CP$ asymmetry via a simultaneous unbinned extended maximum likelihood fit of \dz and \adz samples to the invariant mass 
of the \dz candidates and the cosine of the helicity angle $\theta_H$. The latter is the angle between the momenta of the \dz and the $\pi^+$, $K^+$, or $K^-$ 
in the rest frame of the \rhoz, $\phi$, or \aksz, respectively. By angular momentum conservation, the signal $\cos\theta_H$ distribution depicts a $1-\cos^2\theta_H$ 
dependence; no background contribution is expected to exhibit a similar shape. For the \rhoz and \aksz modes, we restrict the helicity angle range 
to $-0.8<\cos\theta_H<0.4$ to suppress backgrounds that peak at the edges of the distribution. For the $\phi$ mode, where the background levels are lower 
overall, the entire $\cos\theta_H$ range is used. The \dz candidate mass is restricted to  $\SI{1.67}{\gevcc}<M(\dz)<\SI{2.06}{\gevcc}$ for all three signal channels. 

The invariant mass distribution of signal events is modeled with a Crystal-Ball probability density function~\cite{Skwarnicki:1986xj} (PDF) for the 
\rhoz and $\phi$ modes, and with the sum of a Crystal-Ball and two Gaussians for the \aksz mode. To take into account possible differences between MC and data, 
a free offset and scale factor are implemented for the mean and width of the \aksz PDF, respectively. The obtained values are applied to the other two modes.

The $\pi^0$- and $\eta$-type background $M(\dz)$ distributions are described with a pure Crystal-Ball or the sum of either a Crystal-Ball or 
logarithmic Gaussian~\cite{Ikeda:1999aq} and up to two additional Gaussians. For the \rhoz mode, the $\pi^0$-type backgrounds are 
$\rho^0 \pi^0$, $\rho^{\pm}\pi^{\mp}$ and $K^- \rho^+$ with the kaon being misidentified as pion. For the $\phi$ mode, the only $\pi^0$-type background is the decay 
$\dz \to \phi \pi^0$. For the \aksz mode, the $\pi^0$- and $\eta$-type backgrounds are the decays $\dz \to\aksz \pi^0,~K^-\rho^+,~K{}_{0}^{*}(1430)^{-}\pi^+,~K{}^{*-}\pi^+,$ nonresonant $K^- \pi^+ \pi^0$, $\aksz \eta$ and nonresonant $K^- \pi^+ \eta$.  In all three signal modes, the `other-$D^0$' background 
comprises all other decays wherein the \dz is reconstructed from the majority of daughter particles. In the \rhoz(\aksz) mode, there are two additional 
small backgrounds: $\pi^+\pi^-(K^-\pi^+)$ with the photon being emitted as final state radiation (FSR), and $K^-\rho^+$ with the photon arising from the 
radiative decay of the charged $\rho$ meson. As there are no missing particles, these decays exhibit the same $M(\dz)$ distribution as the signal decays. 
We jointly denote them as irreducible background. Their yields are fixed to MC expectations and the known branching fractions~\cite{Agashe:2014kda}. The remaining combinatorial background is parametrized in $M(\dz)$ with an exponential function in the $\phi$ mode and a 
second-order Chebyshev polynomial in the \rhoz and \aksz modes. All parameters describing the combinatorial background are allowed to vary in the fit. Possible correlations among 
the fit variables are negligible, except for the $\aksz \pi^0$ and $K^-\rho^+$ backgrounds in the \aksz mode that are accomodated with an additional Gaussian in the mass PDF whose relative contribution is a function of $\cos\theta_H$.

The $M(\dz)$ PDF shape for the $\pi^0 (\eta)$-type background, obtained from MC samples, is calibrated using the forbidden decay $\dz \to K{}_S^0 \gamma$, 
which yields mostly background from  $\dz \to K{}_S^0 \pi^0$ and $\dz \to K_S^0 \eta$. The same PID criteria as for signal decays are applied, 
along with the $q$ and $p_{\mathrm{CMS}}(\dsp)$ requirements as determined for the $\phi$ mode. The $K{}_S^0 \to \pi^+ \pi^-$ candidates in a 
\SI{\pm9}{\mevcc} window around the nominal mass are accepted. To calibrate the distribution, the simulated shape is smeared with a Gaussian function 
of width \SI{7\pm1}{\mevcc} and an offset \SI{-1.33\pm0.25}{\mevcc}.

The  $\cos\theta_{H}$ signal distribution is parametrized as $1-\cos^2\theta_{H}$ for all three modes. For the $V \pi^0$ and $V \eta$ ($V=\rhoz,\, \phi,\, \aksz$) categories,  
the shape is close to $\cos^2\theta_{H}$ and described with a second- (\rhoz and $\phi$  mode) or third-order (\aksz mode)  Chebyshev polynomial. 
In the $\phi$ mode, a linear term in $\cos\theta_{H}$ is added with a free coefficient to take into account possible interference between resonant 
and nonresonant amplitudes. For other background categories, the distributions are modeled using suitable PDFs based on MC predictions.

Apart from normalizations, the asymmetries $A_{\mathrm{raw}}$ of signal and background modes are left free in the fit. All PDF shapes are fixed to MC values, 
unless previously stated otherwise.

In the \aksz mode, the yields (and $A_{\mathrm{raw}}$) of certain backgrounds that contain a small number of events (one or two orders of magnitude less than signal) 
are fixed: $K{}_{0}^{*}(1430)^{-}\pi^+,~K{}^{*-}\pi^+$, and the `other-$D^0$' background. The same is done for backgrounds with a photon from  FSR or  
radiative $\rho$ decay in the \rhoz and \aksz modes. All fixed yields are scaled by the ratio between reconstructed signal events in data and simulation of 
the normalization modes. We impose an additional constraint in the \aksz mode by assigning two common $A_{\mathrm{raw}}$ variables to 
$\pi^0$- and $\eta$-type backgrounds, respectively. Since all are Cabibbo-favored decays, \Acp is expected to be zero, while other asymmetries 
contributing to $A_{\mathrm{raw}}$ are  the same for decays with the same final-state particles.

\begin{figure}
\centering 
\includegraphics[width=0.42\textwidth]{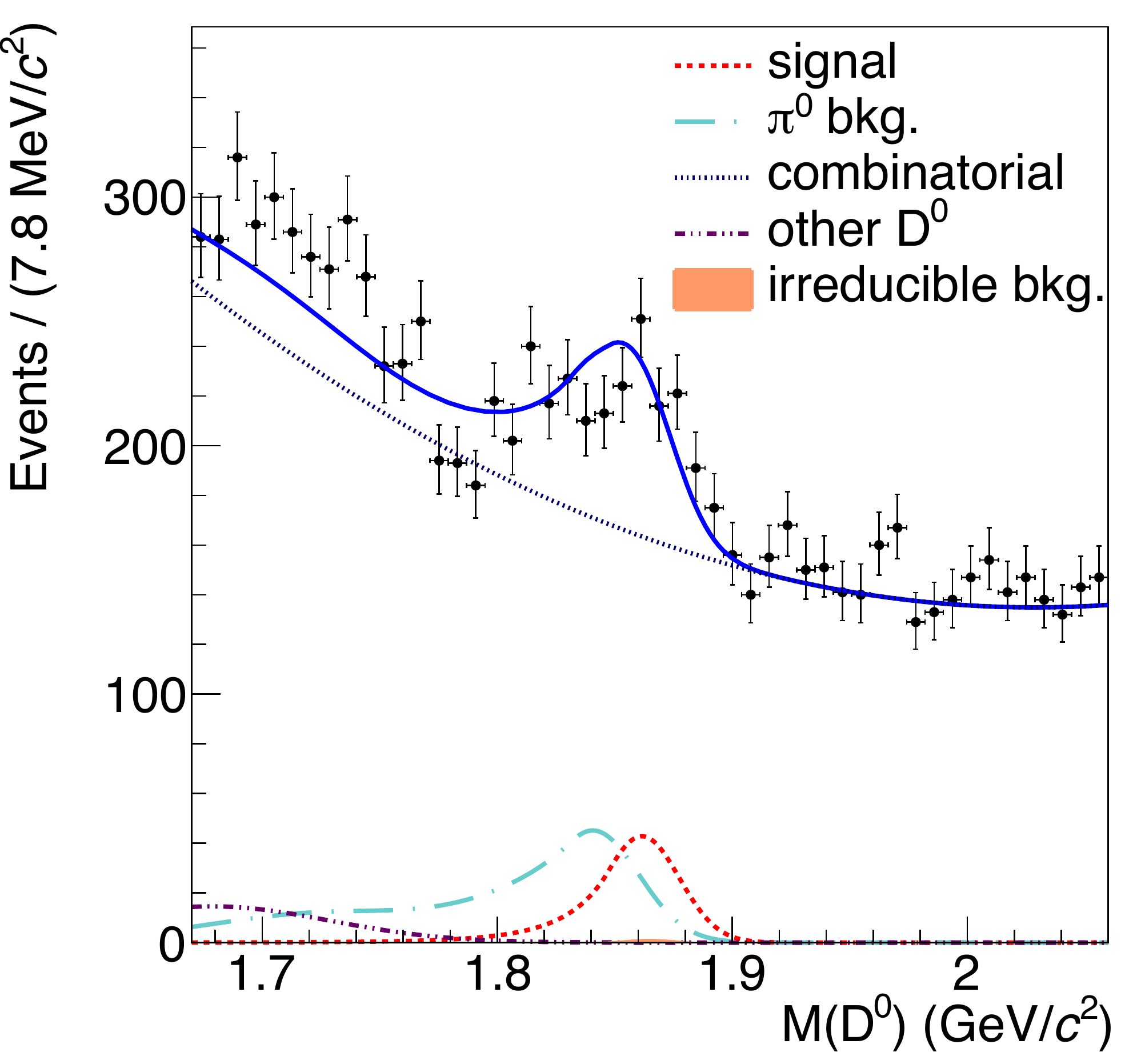}
\includegraphics[width=0.42\textwidth]{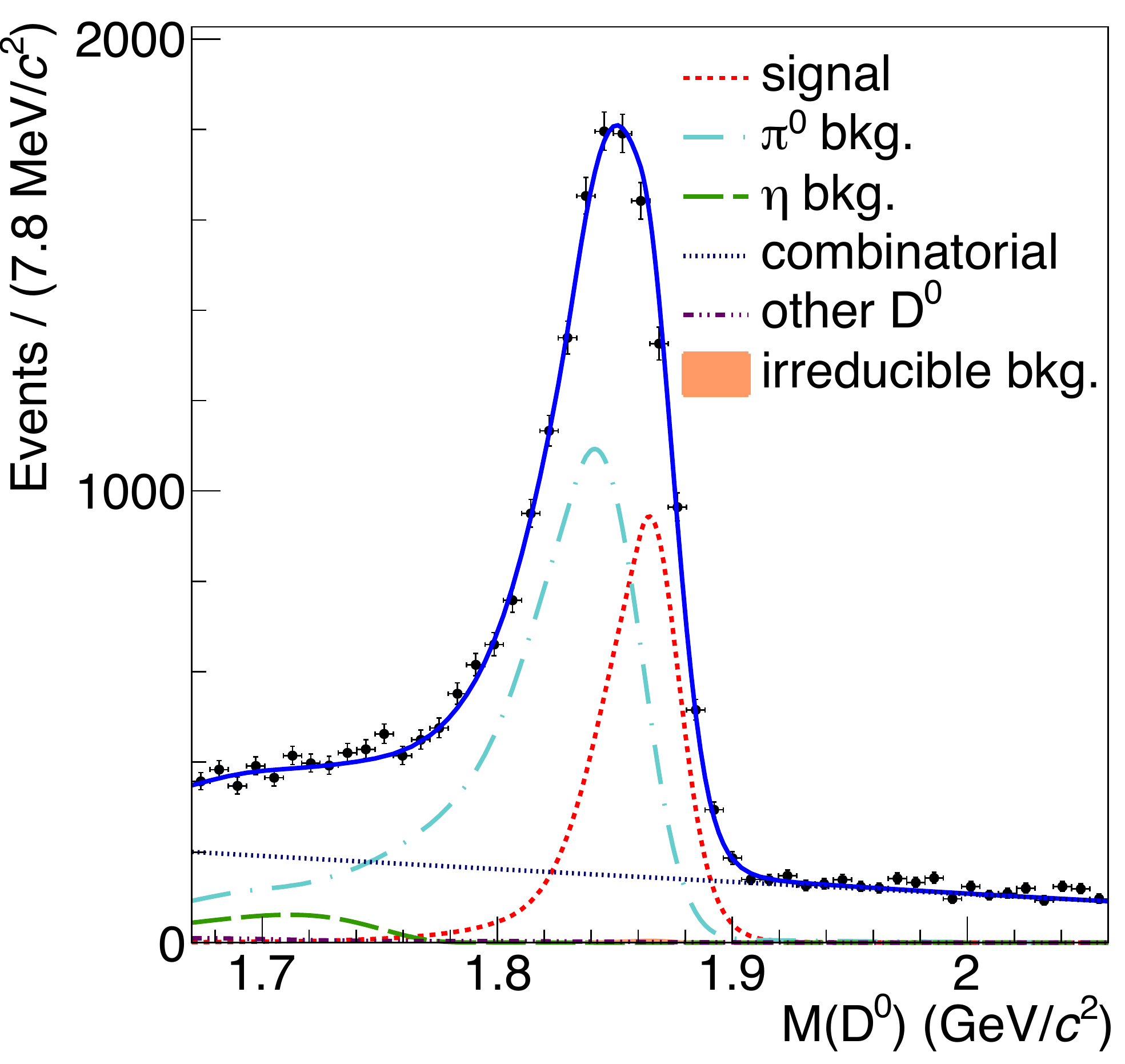}\\
\includegraphics[width=0.42\textwidth]{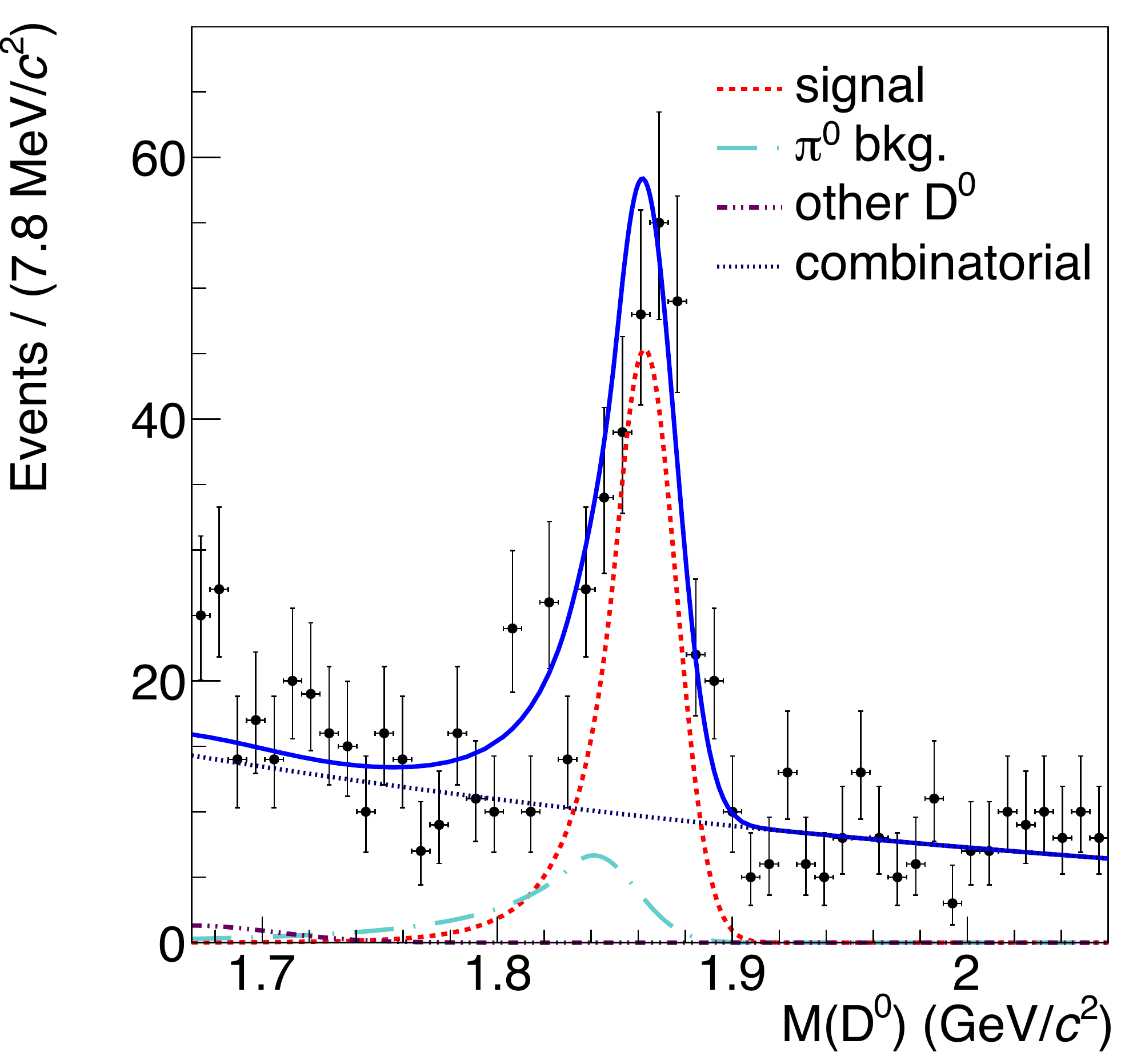}
\includegraphics[width=0.42\textwidth]{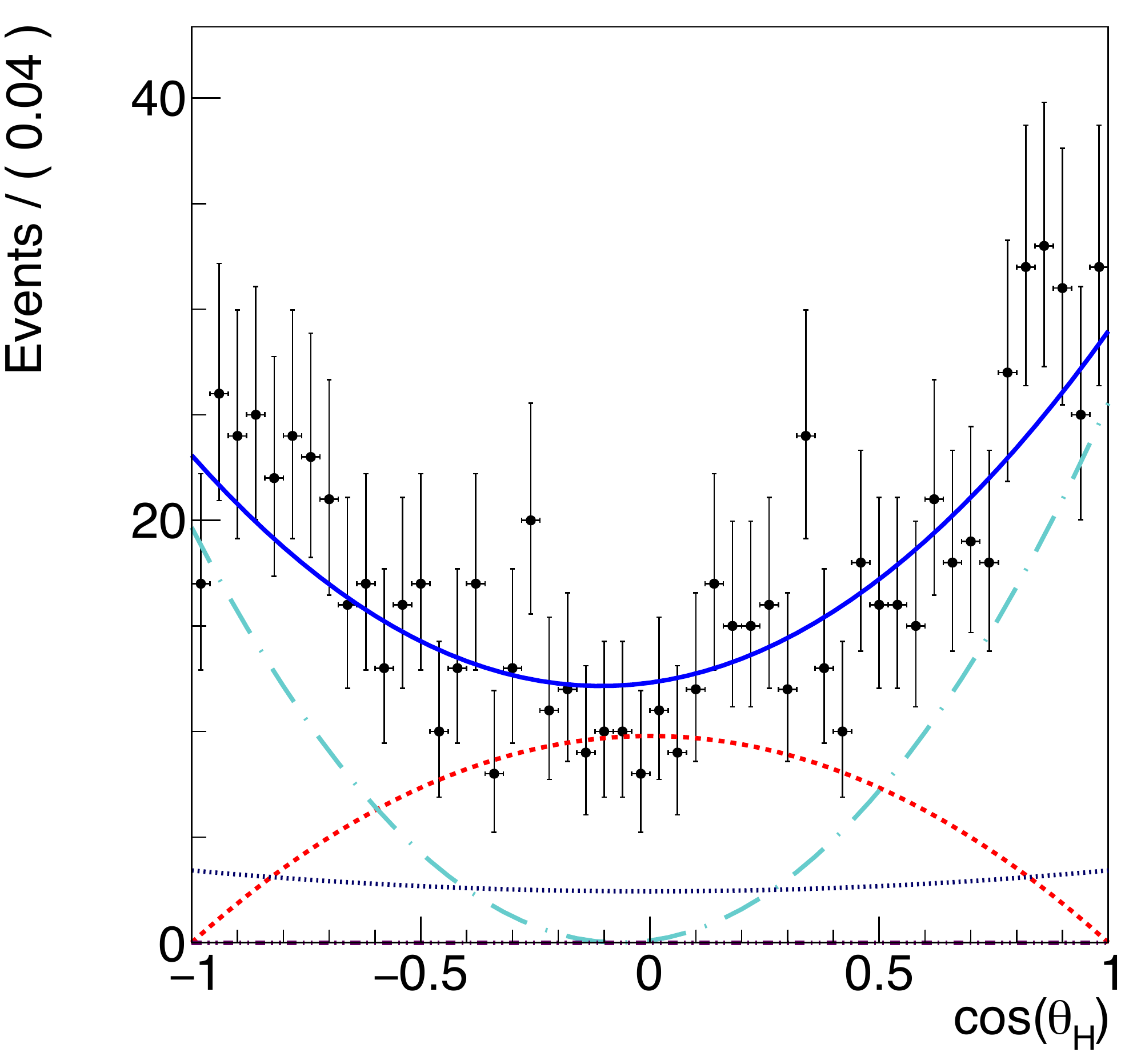}
\caption{Top two panels are signal-enhanced projections of the combined $M(\dz)$ distribution for \dztorg (left) and $\aksz\gamma$ (right). 
Bottom two panels are the signal-enhanced $M(\dz)$ (left) and $\cos\theta_H$ (right) distributions for \dztopg. Fit results are superimposed, 
with the fit components identified in the panel legend.}
\label{fig}
\end{figure}

Fig.~\ref{fig} shows the signal-enhanced $M(\dz)$ projections of the combined sample in the region $-0.3 < \cos\theta_H < 0.3$ for all three signal modes, 
as well as the signal-enhanced $\cos\theta_H$ projection in the $\SI{1.85}{\gevcc}<M(\dz)<\SI{1.88}{\gevcc}$ region for the $\phi\gamma$ mode~\cite{supplemental}. 
The obtained signal yields and raw asymmetries are listed in Table~\ref{tab:num}, along with reconstruction efficiencies. The background raw asymmetries are consistent with zero.

\begin{table}
\caption{Efficiencies, extracted yields and $A_{\mathrm{raw}}$  values for all signal and normalization modes. The uncertainties are statistical. \label{tab:num}}
\begin{tabular}{l|ccc}
\hline \hline
				& Efficiency [\%] 	&	Yield		& $A_{\mathrm{raw}}$   \\ \hline
$\rho^0 \gamma$			&$6.77 \pm 0.09$	&\num{500\pm85}		&$+0.064 \pm 0.152$\\				
$\phi \gamma$ 			&$9.77 \pm 0.10$	&\num{524\pm35}		&\num{-0.091\pm0.066}	\\
$\aksz \gamma$			&	$7.81 \pm 0.03$	&\num{9104\pm396}	&\num{-0.002\pm0.020} \\
\hline
$\pi^+\pi^-$			&$21.4 \pm 0.12$	&\num{1.28\pm0.01e05}	&\num{8.1\pm3.0e-03}\\
$K^+K^-$ 			&$22.7 \pm 0.12$	&\num{3.62\pm0.01e05} 	&\num{2.2\pm1.7e-03}\\
$K^-\pi^+$			&$27.0 \pm 0.13$	&\num{4.02\pm0.02e6}	&\num{1.3\pm0.5e-03}\\
\hline
\hline
\end{tabular}
\end{table}

The analysis of the normalization modes relies on the previous analysis by Belle~\cite{Staric:2008rx}. The same selection criteria as for signal modes 
for PID, vertex fit, $q$ and $p_{\mathrm{CMS}}(\dsp)$ are applied. The signal yield is extracted by subtracting the background in a signal window  of $M(\dz)$, 
where the background is estimated from a symmetrical upper and lower sideband.  
The signal window and sidebands for the $\pi^+\pi^-$   mode are \SI{\pm15}{ \mevcc} and $\pm$(20-35){ \mevcc} around the nominal value~\cite{Agashe:2014kda}, respectively. 
For the $K^+K^-$ mode, the signal window is \SI{\pm14}{ \mevcc} and sidebands are $\pm$(31-45){ \mevcc}, whereas for the $K^-\pi^+$  mode, 
the signal window is \SI{\pm16.2}{ \mevcc} and sidebands are $\pm$(28.8-45.0){ \mevcc}. The obtained signal yields and raw asymmetries are also listed in Table~\ref{tab:num}. 

The systematic uncertainties are listed in Table~\ref{tab:sys}. All uncertainties are simultaneously estimated for \Br and \Acp, unless stated otherwise. There are two main sources: those due to the selection criteria and those arising from 
the signal extraction method, both for signal and normalization modes. Some of the uncertainties from the first group cancel if they are common to the 
signal and respective normalization mode, such as those related to  PID, vertex fit, and the requirement on $p_{\mathrm{CMS}}(\dsp)$. A 2.2\% uncertainty 
is ascribed to photon reconstruction efficiency \cite{Nisar:2015gvd}. Due to the presence of the photon in the signal modes, the resolution of the $q$ distribution 
is worse than in the normalization modes. Thus, the related uncertainties cannot be assumed to cancel completely. We separately estimate the uncertainty due 
to the $q$ requirement using the control channel $\dz \to \aksz \pi^0$. For both MC and data, the efficiency is estimated by calculating the ratio $R$ of the signal yield, 
extracted with and without the requirement on $q$. Then, the double ratio ${R_{\mathrm{MC}}}/{R_{\mathrm{data}}}$ is calculated to assess the possible 
difference between simulation and data. We obtain ${R_{\mathrm{MC}}}/{R_{\mathrm{data}}}(q) = 1.0100 \pm 0.0016$. We do not correct the efficiency by 
the central value; instead, we assign a systematic uncertainty of 1.16\%.

The double-ratio method is also used to estimate the uncertainty due to the $\pi^0$-veto requirement on the control channel $\dz \to K_S^0 \pi^0$. 
The veto is calculated by pairing the first daughter photon (the more energetic one) of the $\pi^0$ with all others, but for the second daughter. 
The ratio $R$ of so-discarded events is calculated for MC and data,  with all other selection criteria applied. The obtained double ratio is 
${R_{\mathrm{MC}}}/{R_{\mathrm{data}}}(\pi^0 \, \mathrm{veto})= 1.002 \pm 0.005$. The error directly 
translates to the systematic uncertainty of the efficiency. 

The systematic uncertainties due to the ${E_{9}}/{E_{25}}$ and $E_{\gamma}$ requirements are estimated on the \aksz mode by repeating the 
fit without any constraint on the variable in question. The systematic error is the difference between the central value of the ratio 
${N_{\mathrm{sig}}}/{\varepsilon_{\mathrm{sig}}}$ from this fit and that of the nominal fit. The obtained uncertainties are 0.23\% for 
${E_{9}}/{E_{25}}$ and 1.15\% for $E_{\gamma}$.

The systematic uncertainties due to the requirement on the mass of the vector meson are estimated using the mass distribution, 
modeled with a relativistic Breit-Wigner function. In the signal window, we compare the integrals of the nominal function and the same modified 
by the uncertainties on the central value and width. The obtained uncertainties are 0.2\% for the \rhoz mode, 0.1\% for the $\phi$ mode, and 
1.7\% for the \aksz mode. All uncertainties described above are summed in quadrature and the final value 
is listed as `Efficiency' in Table~\ref{tab:sys}. They affect only the branching fraction, as they cancel in Eq.~\ref{eq:arawdef}.

For the fit procedure, a systematic uncertainty must be ascribed to every parameter that is determined and fixed to MC values 
but might differ in data. The fit procedure is repeated with each parameter varied by its uncertainty on the positive and negative sides. 
The larger deviation from the nominal branching fraction or \Acp value is taken as the double-sided systematic error and these are summed in quadrature
for all parameters. An uncertainty is assigned to the calibration offset and width of the $\pi^0$-type backgrounds. For the $\phi$ and $\rho^0$ modes, 
the uncertainty is calculated for the width scale factor (and offset) of the signal $M(\dz)$ PDF and $\pi^0$-type background varied simultaneously. 
All these quadratically summed uncertainties are listed as `Fit parametrization' in Table~\ref{tab:sys}.

The values of the fixed yields of some backgrounds in the $\rho^0$ and $\aksz$ mode are varied according to the uncertainties of the respective branching fractions~\cite{Agashe:2014kda}. For the category with the FSR photon, a 20\% variation is used~\cite{Benayoun:1999hm}. As the 
branching fractions contributing to the `other-$D^0$' background in the $\aksz$ mode are unknown, we apply the largest variation from among other categories. 
The quadratically summed uncertainty is listed as `Background normalization' in Table~\ref{tab:sys}.

For the normalization modes, the procedure is repeated with shifted sidebands, starting from \SI{\pm25}{\mevcc} from the nominal $m(\dz)$ value. 
The statistical error from sideband subtraction is taken into account. Since possible differences in the signal shape between simulation 
and data could also affect the signal yield, a similar procedure as for the calibration of the $\pi^0$ background is performed. 
A systematic uncertainty is assigned for the case when the MC shape is smeared by a Gaussian of width \SI{1.6}{\mevcc}. 
All uncertainties arising from normalization modes are summed in quadrature and listed as `Normalization mode' in Table~\ref{tab:sys}.

Finally, an uncertainty is assigned by varying the nominal values of the branching fractions and \Acp of the normalization modes and 
vector meson sub-decay modes by their respective uncertainties.

\begin{table}
\caption{Systematic uncertainties for all three signal modes. \label{tab:sys}}
\begin{tabular}{l|ccc|ccc}
\hline \hline
				&  \multicolumn{3}{c|}{$\sigma$(\Br)/\Br [\%]} 		&  \multicolumn{3}{c}{$A_{CP}$ [$\times 10^{-3}$]} \\
				& $\phi$ 		& $\aksz$		& $\rho^0$		& $\phi$ 		& $\aksz$		& $\rho^0$ \\ \hline
Efficiency			&	2.8		&	3.3		&	2.8		&	--		&	--		&	--	\\
Fit parametrization		&	1.0		&	2.8		&	2.3		&	0.1		&	0.4		&	5.3\\ 
Background normalization 	&	--		&	0.3		&	0.6		&	--		&	0.2		&	0.5\\ 
Normalization mode		&	0.0		&	0.0		&	0.1		&	0.5		&	0.0		&	0.3\\ 
External {\Br} and {\Acp} 	&	2.0		&	1.0		&	1.8		&	1.2		&	0.0		&	1.5\\
\hline
Total				&	3.6		&	4.5		&	4.1		&	1.3		& 	 0.4		& 	5.5\\
\hline
\hline
\end{tabular}
\end{table}

We have conducted a measurement of the branching fraction and \Acp in three radiative charm decays 
$\dztorg$, $\phi\gamma$, and $\aksz \gamma$ using the full dataset recorded by the Belle experiment. We report the first observation of  \dztorg 
with a significance of 5.5$\sigma$, including systematic uncertainties. The significance is calculated as 
$\sqrt{-2\ln( {\mathcal{L}_0}/{ \mathcal{L_{\mathrm{max}}}})}$, where $\mathcal{L}_0$ is the likelihood value with the signal yield fixed to zero 
and $\mathcal{L_{\mathrm{max}}}$ is that of the nominal fit. The systematic uncertainties are included by convolving the statistical likelihood function 
with a Gaussian of width equal to the systematic uncertainty that affects the signal yield. The measured ratios of branching fractions 
to their normalization modes are $(1.25 \pm 0.21 \pm 0.05) \times 10^{-2}$, $(6.88 \pm 0.47 \pm 0.21) \times 10^{-3}$ and 
$(1.19 \pm 0.05 \pm 0.05) \times 10^{-2}$ for \dztorg, $\phi \gamma$, and $\aksz \gamma$, respectively. The first uncertainty is statistical and 
the second systematic. Using world-average values for the normalization modes~\cite{Agashe:2014kda}, we obtain

\begin{align*}
\brdztorg&=(1.77 \pm 0.30 \pm 0.07) \times 10^{-5} ,\\
\brdztopg&=(2.76 \pm 0.19 \pm 0.10) \times 10^{-5},\\
\brdztokg&=(4.66 \pm 0.21 \pm 0.21) \times 10^{-4}.
\end{align*}
For the \rhoz mode, the obtained value is considerably larger than theoretical expectations~\cite{Khodjamirian:1995uc, Fajfer:1998dv}. The result 
of the $\phi$ mode is improved compared to the previous determinations by Belle and \babar{}, and is consistent with the 
world average value~\cite{Agashe:2014kda}. Our branching fraction of the \aksz mode is 3.3$\sigma$ above the \babar{} measurement~\cite{Aubert:2008ai}. 
Both $\phi$ and \aksz results agree with the latest theoretical calculations~\cite{Fajfer:2015zea}.

We also report the first measurement of \Acp in these decays. The values, obtained from Eq.~\ref{eq:acpcalc}:

\begin{align*}
 \acpdztorg&=+0.056 \pm 0.152 \pm 0.006,\\
\acpdztopg&=-0.094 \pm 0.066 \pm 0.001,\\
\acpdztokg&=-0.003 \pm 0.020 \pm 0.000,\\
\end{align*}
 are consistent with no $CP$ violation. Since the uncertainty is statistically dominated, the sensitivity can be greatly enhanced at the upcoming Belle II experiment~\cite{Abe:2010gxa}.


\begin{acknowledgments}
We thank the KEKB group for excellent operation of the
accelerator; the KEK cryogenics group for efficient solenoid
operations; and the KEK computer group, the NII, and 
PNNL/EMSL for valuable computing and SINET4 network support.  
We acknowledge support from MEXT, JSPS and Nagoya's TLPRC (Japan);
ARC (Australia); FWF (Austria); NSFC and CCEPP (China); 
MSMT (Czechia); CZF, DFG, EXC153, and VS (Germany);
DST (India); INFN (Italy); 
MOE, MSIP, NRF, BK21Plus, WCU and RSRI  (Korea);
MNiSW and NCN (Poland); MES and RFAAE (Russia); ARRS (Slovenia);
IKERBASQUE and UPV/EHU (Spain); 
SNSF (Switzerland); MOE and MOST (Taiwan); and DOE and NSF (USA).
\end{acknowledgments}

\bibliography{bibliography}
\end{document}